# Classification of the sign of the critical Casimir force in two dimensional systems at asymptotically large separations

M. A. Rajabpour[1]

[1] *Instituto de Física, Universidade Federal Fluminense,*
*Av. Gal. Milton Tavares de Souza s/n, Gragoatá, 24210-346, Niterói, RJ, Brazil*
(Dated: August 27, 2018)

We classify the sign of the critical Casimir force between two finite objects separated by a large distance in the two dimensional systems that can be described by conformal field theory (CFT). In particular, we show that as far as the smallest scaling dimension present in the spectrum of the system is smaller than one, the sign of the force is independent of the shape of the objects and can be determined by the elements of the modular $S$-matrix of the CFT. The provided formula for the sign of the force indicates that the force is always attractive for equal boundary conditions independent of the shape of the objects. However, different boundary conditions can lead to attractive or repulsive forces. Using the derived formula, we prove the known results regarding the Ising model and the free bosons. As new examples, we give detailed results regarding the Q=3-states Potts model and the compactified bosons. In particular, for the latter model we show that Dirichlet boundary condition does not always lead to an attractive force.

PACS numbers:

Two neutral objects placed near each other at distances of a few micrometers interact with each other via Casimir force [1]. The Casimir force has been the subject of intense studies in the last eighty years in many different areas of research, for review see [2, 3]. Although the first studies were mostly focused on the Casimir force in the quantum electrodynamic (QED), it was soon realized that the same effect can also arise for two objects embedded in any kind of a critical medium [4, 5]. The Casimir effect has been also studied experimentally from the very beginning [6, 7], however, the main breakthrough in precise measurements just happened in more recent times, for QED Casimir effect see [8–10] and for critical fluctuation-induced forces see [11–13].

Determining the sign of the Casimir force has been the subject of interest from the very beginning. In [14] it was shown that in dielectric systems both attractive and repulsive forces can arise, for more recent studies on the conditions of the repulsive forces, see [15]. Apart from theoretical interest characterizing the sign of the Casimir force can be very important in nano, and micro-sciences to avoid the stiction. Although now it became an old subject, it is quite interesting that the problem of the classification of the sign of the Casimir force is still an open problem, for recent advances see [16–24]. One of the main difficulties is the shape dependence of the sign of the Casimir force in the three dimensional systems. However, remarkably, there is a theorem for mirror symmetric objects subject to Dirichlet boundary conditions in arbitrary dimensions which guarantees that the force is always attractive in those systems [16, 17]. There are also some general results regarding the stability of the objects acting through the Casimir force, see [21] and references therein. Although most of the effort in characterizing the sign of the Casimir force has been in three dimensional systems the recent experiments on lipid mixtures composing biological membranes [25, 26] motivated the necessity of more thorough investigation of the Casimir effect in two dimensional systems. For example, in [27] it was argued that Casimir-like forces between membrane bound inclusions can arise in two dimensions and they can be studied in the realm of the two dimensional Ising model. Casimir Force in the two dimensional membranes has been already studied for a long time, see for example [28–30] and references therein. However, recent studies [27, 31] based on boundary conformal field theory opened a unified way to study the critical Casimir force in two dimensions for arbitrary objects and generic universality classes. The basic idea is that if the two objects embedded in the medium induce conformal boundary conditions one can map the system with two holes to an annulus and since the Casimir energy on the annulus is known one can extract exact generic formulas for the Casimir force between two arbitrary objects which depends on the conformal map which takes the system with two holes to an annulus. The same formulas were recently used to calculate the entanglement entropy [32] and the formation probabilities [33] in the one dimensional quantum chains. In this letter, we will use this line of work to classify the sign of the Casimir force between two finite objects embedded in a far distance. It is worth mentioning that similar idea has been already used in [22] to study the sign of the Casimir force between two infinite size systems which in that study it was concluded that the sign is usually dependent on the shape of the objects. However, in this letter, we prove that in a vast majority of the cases the sign of the Casimir force between two finite objects is independent of the shape of the objects and can be actually classified based on the type of the boundary conditions induced by the objects. We first review the results of [27, 31] and introduce few useful formulas regarding the boundary conformal field theory on the annulus. Then we derive the Casimir force between the objects with a sign which explicitly depends on the elements of the modular $S$ matrix of the underlying CFT. Using the formula one can find the sign of the Casimir force practically for all the models that the boundary CFT is known. In particular, we derive the already predicted results regarding the Ising model and the free bosons. Then as new examples we provide the sign of the Casimir force for different boundary conditions in the Q=3-states Potts model and the compactified bosons.

As we already mentioned in the introduction, embedding two objects in the medium of a two-dimensional critical system is like enforcing boundary conditions on the domains of the objects. We call the regions occupied by the objects $D_1$ and $D_2$ and the distance between their origins $|z_{12}|$. To calculate the Casimir force between the two objects, first one needs to calculate the free energy necessary to bring the two objects to the distance $|z_{12}|$. If the boundary conditions on the boundaries of $D_1$ and $D_2$ respect the conformal symmetry of the bulk then one can use the methods of CFT to calculate the Casimir free energy and ultimately the Casimir force. The idea goes as follows [27, 31]: The free energy of most of the well-known CFTs are calculated on the annulus [34], we call it $\mathcal{F}^{an}$. One the other hand, because of the conformal symmetry one can map any geometry with two boundaries to an annulus with inner and outer radius $e^{-h}$ and $1$ with a conformal map $w(z)$, see figure 1. The conformal mapping also produces a contribution to the free energy which we call the geometric part of the Casimir energy $\mathcal{F}^{ge}$, see [31]. Finally one can write the Casimir force as:

$$F = \frac{F_x + iF_y}{2} = -\partial_{z_{12}} \mathcal{F}^{an} - \partial_{z_{12}} \mathcal{F}^{ge}; \quad (1)$$

where the geometric part is just dependent on the conformal map $w(z)$ and can be written as:

$$\partial_{z_{12}} \mathcal{F}^{ge} = \frac{ic}{24\pi} \oint_{D_2} \{w, z\} dz, \quad (2)$$

where $c$ is the central charge of the system and the contour of the integral is around the domain $D_2$. It was shown in [31] that when the two objects are far from each other, the contribution of the geometric part to the Casimir force is

$$F^{ge} = -\frac{(Q_{1,1}^2 + Q_{1,2})(Q_{2,1}^2 + Q_{2,2})}{z_{12}^5} + \mathcal{O}(z_{12}^{-6}); \quad (3)$$

where the coefficients $Q_{i,j}$ appear in the bipolar expansion of the conformal map $w(z)$. Note that since we are not aware of possible restrictions on the values of the coefficients $Q_{i,j}$, which are dependent on the geometry of the embedded objects, we assume that the $F^{ge}$ can be positive or negative [35]. However, as we discuss now the main contribution to the Casimir force comes from the annulus part and can be classified into different universality classes. It is worth mentioning that in some interesting cases, such as two discs, the corresponding conformal map is just a Möbius transformation and consequently $F^{ge} = 0$. For these geometries the sign of the Casimir force can be determined using just the annulus part.

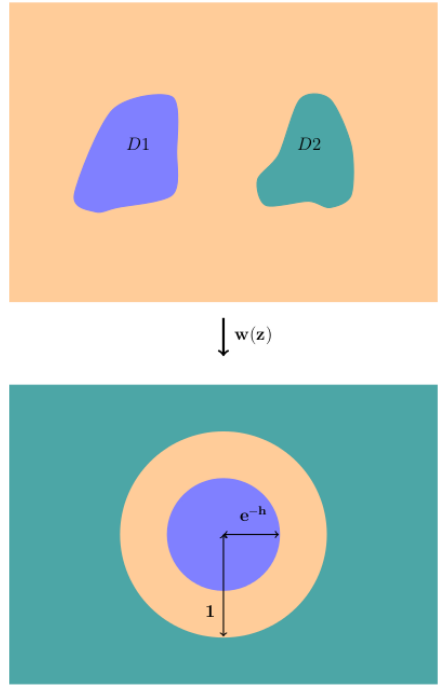

FIG. 1: (Color online) The conformal map $w(z)$ takes the whole plane minus the domains $D_1$ and $D_2$ to an annulus.

The free energy on the annulus with boundary conditions $A$ and $B$ on the two boundaries can be written with respect to the Virasoro characters of the CFT, however, it is much better to write the series form as follows [34, 36]:

$$\mathcal{F}^{an}_{AB}(\tilde{q}) = \\ -\ln[\tilde{q}^{-c/24}(b_0^A b_0^B + \sum_j b_j^A b_j^B \tilde{q}^{\Delta_j})] + c\frac{h}{12}, \quad (4)$$

where $\tilde{q} = e^{-2h}$ and $b_j^A = \langle A|j\rangle\rangle$ and $b_j^B = \langle\langle j|B\rangle$. $|A(B)\rangle$ and $|j\rangle\rangle$ are Cardy and Ishibashi states respectively. The coefficients $b_j^{A(B)}$ are related to the fusion coefficients $n_j^{AB}$ of the corresponding conformal field theory with the Verlinde formula $n_j^{AB} = \sum_{j'} S_j^{j'} b_{j'}^A b_{j'}^B$, where $S_j^{j'}$ is the element of the modular matrix $S$, see [36]. Note that all the coefficients can be written with respect to the elements of the modular matrix, for example, $b_j^0 = \langle\langle j|0\rangle = (S_0^j)^{1/2}$ and for $\Delta \neq 0$ we have $b_j^\Delta = \frac{S_\Delta^j}{(S_0^j)^{1/2}}$. All the coefficients are well-known for most of the rational CFTs [37]. Although the fusion coefficients $n_j^{AB}$ are all non-negative integers, the other coefficients can be positive or negative real numbers. Excep-

tionally $b_0^{A(B)}$ and $b_j^0$ are always non-negative real numbers. Finally $\Delta_j$'s in the equation (4) are the conformal weights of the bulk operators propagating around the annulus. Note that in the sum, we have all the highest weights and their descendants.

Since for the large separations of the two objects we have $h \to 2\ln|z_{12}|$ one can write

$$\mathcal{F}_{AB}^{an}(z_{12}) \to -\ln[b_0^A b_0^B] - \frac{b_1^A b_1^B}{b_0^A b_0^B} \frac{1}{|z_{12}|^{4\Delta_1}}, \quad (5)$$

where $\Delta_1$ is the smallest scaling dimension present in the spectrum of the system. The first term is the Affleck-Ludwig boundary entropy and the second term is the leading decaying term in the annulus part of the free energy. Although the power-law decay $\frac{1}{|z_{12}|^{4\Delta_1}}$ has been already predicted for a long time, see [38], deriving the coefficient of the decaying term is the main result of the current work. As far as $\Delta_1 < 1$, which is the most common case (see Appendix A), the main contribution to the Casimir force comes from the annulus part of the free energy. In other words the Casimir force of far away objects decays as $F \to -\frac{b_1^A b_1^B}{b_0^A b_0^B} \frac{1}{|z_{12}|^{4\Delta_1+1}}$. Then one can determine the sign of the force $\boldsymbol{sn}$ by

$$\boldsymbol{sn} = -\frac{b_1^A b_1^B}{|b_1^A b_1^B|}. \quad (6)$$

The above equation is our main result and although it has a very simple form it can be used to classify the sign of the Casimir force for a wide variety of critical systems. One of the immediate consequences of the above formula is that for the same boundary conditions on the domains $D_1$ and $D_2$ the Casimir force is always attractive. Note that this result is independent of the geometry of the objects as far as $\Delta_1 < 1$. For those cases that the main contribution comes from the geometric part of the force the sign can be dependent on the shape of the objects. Since all the coefficients $b_j^{A(B)}$ are known for most of the CFTs, one can use them to find the sign of the Casimir force. In principle, as we will show, for different boundaries the sign can be positive or negative. From now on we focus on three examples, Ising model, Q=3 state Potts model and free bosons. In the case of the Ising model and free bosonic systems, we prove the already known results. For compactified bosonic systems we produce some new surprising results. All the conclusions regarding Q=3 state Potts model are new.

**Ising model** - The model is defined by the Hamiltonian

$$H = -J \sum_{<ij>} s_i s_j, \quad (7)$$

where $s_i = \pm 1$ and the sum is over all the nearest neighbor sites. The possible conformal invariant boundary conditions of the Ising model are known and are called *fixed* and *free*, see the appendix B. Since the fixed boundary condition can be up or down (+ or −), we have four different possibilities for the partition functions on the annulus: a) (++); b) (+−); c) (+f) and d) (ff). The first and the last cases, as we discussed before, lead to attractive forces. Using the $b$ coefficients listed in the appendix B for the sign of the Casimir force, i.e. $\boldsymbol{sn}$ we have

$$\boldsymbol{sn}(++) < 0 \quad (8)$$
$$\boldsymbol{sn}(+-) > 0 \quad (9)$$
$$\boldsymbol{sn}(+f) > 0 \quad (10)$$
$$\boldsymbol{sn}(ff) < 0 \quad (11)$$

The above conclusions are consistent with the known results, see for example [5]. However note that here we prove them independent of the shape of the floating objects.

**Q=3 States Potts model-** The model is defined by the Hamiltonian

$$H = -J \sum_{<ij>} \delta_{\sigma_i \sigma_j}, \quad (12)$$

where $\sigma_i = 1, 2, 3$ and the sum is over all the nearest neighbor sites. Many different boundary conditions are possible for this model. To keep the discussion as simple as possible we confine ourselves to the ones discussed in [34, 42]. Because of the three spin possibilities one can define three fixed boundary conditions as $a$, $b$ and $c$. However, it is also possible to define three mixed boundary conditions $a+b$, $a+c$ and $b+c$. Finally, there is also a free boundary condition. These seven boundary conditions can be paired up in 42 ways, however, just a few of them are independent and the rest can be derived using the $Z_3$ symmetry of the model. Using the boundary conformal field theory of the model (see the appendix B) we summarized the sign of the Casimir force for different conditions in the TABLE I. Note that although there is a strong tendency for repulsion in the non-equal boundary conditions one can not argue that these boundary conditions always lead to repulsive Casimir force.

TABLE I: Sign of the Casimir force in the Q=3 states Potts model with different boundary conditions.

|       | $a$ | $b$ | $c$ | $a+b$ | $a+c$ | $b+c$ | $f$ |
|-------|-----|-----|-----|-------|-------|-------|-----|
| $a$   | −   | +   | +   | −     | −     | +     | +   |
| $b+c$ | +   | −   | −   | +     | +     | −     | −   |
| $f$   | +   | +   | +   | −     | −     | −     | −   |

**Free bosons-** We now discuss the Casimir force for the free bosonic systems with different boundary conditions. We first study the uncompactified boson which has been already the subject of many studies in the past. Then we discuss compactified bosons which show more interesting features. The free bosonic systems are defined by the action

$$S = \frac{1}{2} \int dx_1 dx_2 [(\partial_1 \phi)^2 + (\partial_2 \phi)^2]. \quad (13)$$

The boundary conformal field theory of the model is well studied, see for example [39]. There are two possible conformal boundary conditions (Dirichlet and Neumann) which can be paired up in three different ways with the partition functions:

$$Z_{DD}^{an}(\tilde{q}) = e^{-\frac{(\phi_0-\phi_0')^2}{8\pi h}} \frac{e^{-\frac{h}{12}}}{\sqrt{2h}} \frac{1}{\eta(\tilde{q})}, \quad (14)$$

$$Z_{NN}^{an}(\tilde{q}) = \frac{e^{-\frac{h}{12}}}{2\eta(\tilde{q})}, \quad (15)$$

$$Z_{DN}^{an}(\tilde{q}) = \frac{1}{\sqrt{2}} \prod_{n=1}^{\infty} \frac{1}{1+\tilde{q}^n}, \quad (16)$$

where $\phi_0(\phi_0')$ is the value of the field on the boundary and $\eta(\tilde{q}) = \tilde{q}^{1/24} \prod_{n=1}^{\infty}(1-\tilde{q}^n)$. Note that for the $NN$ and the $DN$ the smallest scaling dimension is 1 which means that the annulus and the geometric part of the Casimir force are in the same order. This just means that in these two cases the sign of the Casimir force can be dependent on the shape of the objects and naturally one can not draw universal conclusions. However, for the $DD$ boundary conditions we have

$$F_{DD} \to -\frac{1}{2|z_{12}|\ln|z_{12}|} \quad (17)$$

which is an attractive force [31]. Note that the sign of the force is independent of the value of the field on the boundary. The situation is much more intriguing for the compactified boson which is the scaling limit description of a lot of statistical models such as: the Ashkin-Teller model and the Ising model with a defect [40]. Considering $\phi \equiv \phi + 2\pi r$ the annulus partition functions have the following forms, see for example [40]:

$$Z_{DD}^{an}(\tilde{q}) = \frac{1}{2r\sqrt{\pi}} \frac{e^{-\frac{h}{12}}}{\eta(\tilde{q})} \left(1 + 2\sum_{k>0} \cos\frac{k(\phi_0-\phi_0')}{r} \tilde{q}^{\frac{k^2}{8\pi r^2}}\right), \quad (18)$$

$$Z_{NN}^{an}(\tilde{q}) = r\sqrt{\pi} \frac{e^{-\frac{h}{12}}}{\eta(\tilde{q})} \left(1 + 2\sum_{k>0} \cos[2\pi rk(\tilde{\phi}_0-\tilde{\phi}_0')]\tilde{q}^{\frac{\pi k^2 r^2}{2}}\right), \quad (19)$$

where $\tilde{\phi}_0(\tilde{\phi}_0')$ are the values of the dual fields on the lattice. The partition function of $Z_{DN}^{an}$ is as before. For $DD$ as far as $r^2 > \frac{1}{8\pi}$ the smallest scaling dimension is smaller than one and so the annulus part of the partition function is dominant. In this case the sign of the Casimir force is determined by $-\cos\frac{\phi_0-\phi_0'}{r}$. As it is clear the force is attractive whenever $-\frac{\pi}{2} < \frac{\phi_0-\phi_0'}{r} < \frac{\pi}{2}$, and it is repulsive for $\frac{\pi}{2} < \frac{\phi_0-\phi_0'}{r} < \frac{3\pi}{2}$. For the $NN$, the smallest scaling dimension is smaller than one for $r^2 < \frac{2}{\pi}$. In this regime the sign of the Casimir force can be determined by studying $-\cos[2\pi rk(\phi_0-\phi_0')]$. The sign is negative as far as $-\frac{1}{4} < r(\tilde{\phi}_0-\tilde{\phi}_0') < \frac{1}{4}$ and it is repulsive for those cases that $\frac{1}{4} < r(\tilde{\phi}_0-\tilde{\phi}_0') < \frac{3}{4}$. It is interesting to note that in contrast to the non-compactified boson the sign of the Casimir force is dependent on the difference between the values of the fields (dual fields) on the boundaries and in principle can be adjusted. It will be interesting to check this phenomenon directly in the case of the Ashkin-Teller model or in the Ising model with a defect.

**Discussion and conclusions-** In this paper, we showed that the sign of the Casimir force at asymptotically large separations in most of the two dimensional critical systems can be determined by studying the elements of the modular $S$ matrix of the underlying CFT. Although the force is attractive when the boundary conditions induced by the objects are the same, it is not true that different boundary conditions always repel each other. In some cases like the compactified bosons, one can get an alternative behavior for the sign of the Casimir force by changing the value of the (dual)field on the boundary. This is in contrast to common believe that two Dirichlet boundary conditions always attract each other. We note that our analyses can break down if the two objects are too close to each other. One reason is that the geometric part of the force for short distances starts to be important especially for sharp boundaries. At the same time, the annulus part might also change sign depending on the spectrum of the system. It will be very interesting to find a physical argument based on coulomb gas representation for the conclusions that we presented in this article. It is also important to check some of the predictions of the formula presented here with some numerical techniques.

**Acknowledgment:** We thank M Maghrebi for reading the manuscript and comments. The work of MAR was supported in part by CNPq.

## Appendix A: Conformal weights in minimal models

In this appendix, we briefly discuss the conformal weights of minimal models $\mathcal{M}(p,p')$ with integer $p > p' > 0$. For more details see [41]. The discussion is relevant because our classification is heavily based on having conformal dimension $\Delta < 1$. This condition is the most common case, however, there are some exceptions. In minimal models, the central charge $c$ and conformal weights $\Delta_{r,s}$ are positive rational numbers and are given by the following formulas:

$$c = 1 - 6\frac{(p-p')^2}{pp'}; \tag{A1}$$

$$\Delta_{r,s} = \frac{(pr-p's)^2 - (p-p')^2}{4pp'}, \qquad 1 \leq r \leq p'-1, \quad 1 \leq s \leq p-1. \tag{A2}$$

Most of the famous critical models are examples of the above CFT's. For example, $\mathcal{M}(4,3)$, $\mathcal{M}(5,4)$ and $\mathcal{M}(6,5)$ are the Ising, tri-critical Ising, and $Q = 3$ states Potts model respectively. Note that it is not true that all of the scaling operators with the above conformal dimensions appear in every representation of a given statistical model. For example, in the $Q = 3$ states Potts model studied here just those with the $\Delta < 1$ appear. It is simple to verify that in the $\mathcal{M}(4,3)$ there is no conformal weight which is bigger than one. However, in the $\mathcal{M}(5,4)$ one out of six operators has conformal dimension bigger than one. In the $\mathcal{M}(6,5)$ two out of ten scaling dimensions are bigger than one. Note that in all of the above models, the order parameters have conformal dimensions smaller than one. The situation can be clarified most effectively in the case of the RSOS model, see [41]. The model at one of the critical points can be described by the minimal model $\mathcal{M}(q, q-1)$ with $q > 4$. In this model, for example, there are $q-3$ order parameters with the conformal weights $h_{k+1,k+1} = \frac{(k+1)^2-1}{4q(q-1)}$, where $1 \leq k \leq q-3$. All of these conformal weights are smaller than one. The above discussion shows that most of the conformal dimensions are smaller than one in most of the well-known models, however, in some models one might be able to define a new set of boundary conditions where $\Delta > 1$ and consequently one can not use our method to classify the sign of the Casimir force.

### Appendix B: Boundary conformal field theory of the Ising model and the Q=3 states Potts model

In this appendix, we summarize all the necessary formulas of the boundary conformal field theory of the Ising model and the Q=3 states Potts model. All the details can be found in the original paper by Cardy [34, 36], see also [42, 43].

**Ising model -** We first discuss the Ising model which is the simplest rational CFT with the central charge $c = \frac{1}{2}$. The theory has three primary operators: identity with the conformal weight $\Delta_I = 1$, energy operator $\epsilon$ with weight $\Delta_\epsilon = \frac{1}{2}$ and the spin operator $\sigma$ with $\Delta_\sigma = \frac{1}{16}$. The modular $S$ matrix of the model is

$$S = \begin{pmatrix} \frac{1}{2} & \frac{1}{2} & \frac{1}{\sqrt{2}} \\ \frac{1}{2} & \frac{1}{2} & -\frac{1}{\sqrt{2}} \\ \frac{1}{\sqrt{2}} & -\frac{1}{\sqrt{2}} & 0 \end{pmatrix}; \quad \text{(B1)}$$

where the rows and the columns are labeled by the weights $(0, \frac{1}{2}, \frac{1}{16})$. The Conformal boundary states can be now written with respect to the Ishibashi states as

$$|0\rangle = \frac{1}{\sqrt{2}}|0\rangle\rangle + \frac{1}{\sqrt{2}}|\epsilon\rangle\rangle + \frac{1}{2^{\frac{1}{4}}}|\sigma\rangle\rangle; \quad \text{(B2)}$$

$$|\tfrac{1}{2}\rangle = \frac{1}{\sqrt{2}}|0\rangle\rangle + \frac{1}{\sqrt{2}}|\epsilon\rangle\rangle - \frac{1}{2^{\frac{1}{4}}}|\sigma\rangle\rangle; \quad \text{(B3)}$$

$$|\tfrac{1}{16}\rangle = |0\rangle\rangle - |\epsilon\rangle\rangle. \quad \text{(B4)}$$

The first two states can be identified as the fixed $+$ and the fixed $-$ and the last one is identified with the free boundary condition. Having the above states, one can both read all the coefficients $b_j$ and the nature of the smallest scaling dimension present in the system. For example, for the a) $++$ the smallest scaling dimension is $\frac{1}{16}$ and $b_\sigma^+ = \frac{1}{2^{\frac{1}{4}}}$. b) For $+-$ again the smallest scaling dimension is $\frac{1}{16}$ and $b_\sigma^- = -\frac{1}{2^{\frac{1}{4}}}$. c) For fixed-free the smallest scaling dimension is $\frac{1}{2}$ and $b_\epsilon^+ = \frac{1}{2^{\frac{1}{2}}}$ and $b_\epsilon^f = -1$. d) For free-free the smallest scaling dimension is again $\frac{1}{2}$ with $b_\epsilon^f = -1$.

**Q=3 states Potts model -** We now discuss the boundary conformal field theory of Q=3 states Potts model. Although it is possible to discuss the full structure of the boundary conformal field theory of these model, see [43] to avoid unnecessary complications, we start with those cases that respect the W symmetry of the model. The central charge of the CFT is $c = \frac{4}{5}$ and the primary operators are $(I, \epsilon, \psi, \sigma, \psi^\dagger, \sigma^\dagger)$ with the conformal weights $(0, \frac{2}{5}, \frac{2}{3}, \frac{1}{15}, \frac{2}{3}, \frac{1}{15})$. Note that in the Kac table of the $c = \frac{4}{5}$ minimal model we have at least four other operators $(1,2), (4,2), (2,2), (3,2)$ with the conformal weights $\frac{1}{8}, \frac{13}{8}, \frac{1}{40}$ and $\frac{21}{40}$ that play important roles in the Q=3-states Potts model. The modular $S$ matrix of the model is

$$S = N^2 \begin{pmatrix} s & s & s \\ s & \omega s & \omega^2 s \\ s & \omega^2 s & \omega s \end{pmatrix}; \quad s = \begin{pmatrix} 1 & \lambda^2 \\ \lambda^2 & -1 \end{pmatrix}, \quad \text{(B5)}$$

where $N^4 = \frac{5-\sqrt{5}}{30}$ and $\lambda^2 = \frac{1+\sqrt{5}}{2}$. The conformal boundary states can be written as:

$$|0\rangle = N\Big[|0\rangle\rangle + \lambda|\epsilon\rangle\rangle + |\psi\rangle\rangle + \lambda|\sigma\rangle\rangle + |\psi^\dagger\rangle\rangle + \lambda|\sigma^\dagger\rangle\rangle\Big]; \tag{B6}$$

$$|\tfrac{2}{3}\rangle = N\Big[|0\rangle\rangle + \lambda|\epsilon\rangle\rangle + \omega|\psi\rangle\rangle + \omega\lambda|\sigma\rangle\rangle + \omega^2|\psi^\dagger\rangle\rangle + \omega^2\lambda|\sigma^\dagger\rangle\rangle\Big]; \tag{B7}$$

$$|\tfrac{2}{3}^\dagger\rangle = N\Big[|0\rangle\rangle + \lambda|\epsilon\rangle\rangle + \omega^2|\psi\rangle\rangle + \omega^2\lambda|\sigma\rangle\rangle + \omega|\psi^\dagger\rangle\rangle + \omega\lambda|\sigma^\dagger\rangle\rangle\Big]; \tag{B8}$$

$$|\tfrac{2}{5}\rangle = N\Big[\lambda^2|0\rangle\rangle - \lambda^{-1}|\epsilon\rangle\rangle + \lambda^2|\psi\rangle\rangle - \lambda^{-1}|\sigma\rangle\rangle + \lambda^2|\psi^\dagger\rangle\rangle - \lambda^{-1}|\sigma^\dagger\rangle\rangle\Big]; \tag{B9}$$

$$|\tfrac{1}{15}\rangle = N\Big[\lambda^2|0\rangle\rangle - \lambda^{-1}|\epsilon\rangle\rangle + \omega\lambda^2|\psi\rangle\rangle - \omega\lambda^{-1}|\sigma\rangle\rangle + \omega^2\lambda^2|\psi^\dagger\rangle\rangle - \omega^2\lambda^{-1}|\sigma^\dagger\rangle\rangle\Big]; \tag{B10}$$

$$|\tfrac{1}{15}^\dagger\rangle = N\Big[\lambda^2|0\rangle\rangle - \lambda^{-1}|\epsilon\rangle\rangle + \omega^2\lambda^2|\psi\rangle\rangle - \omega^2\lambda^{-1}|\sigma\rangle\rangle + \omega\lambda^2|\psi^\dagger\rangle\rangle - \omega\lambda^{-1}|\sigma^\dagger\rangle\rangle\Big]; \tag{B11}$$

The first three states can be identified as three fixed states $a$, $b$ and $c$. Here, we also report the corresponding partition functions on the cylinder:

$$Z_{a,a} = \chi_I(q) = \chi_{11}(q) + \chi_{41}(q), \tag{B12}$$
$$Z_{a,b} = \chi_\psi(q) = \chi_{13}(q), \tag{B13}$$
$$Z_{a,c} = \chi_{\psi^\dagger}(q) = \chi_{13}(q), \tag{B14}$$

where $\chi_{I,\psi,\psi^\dagger}(q)$ are the W symmetry characters written with respect to the Virasoro characters. Note that by using $\chi(q) = \sum_j S_i^j \chi_j(\tilde{q})$ one can write all the partition functions with respect to the $\tilde{q}$. Note that although in all of the cases the smallest scaling dimension is $\frac{1}{15}$, the $b$ coefficients are different $b^a_{\frac{1}{15}} = -2b^b_{\frac{1}{15}} = -2b^c_{\frac{1}{15}} = 2N\lambda$. The next three states in the list, i.e. (B9), (B10) and (B11), are related to mixed boundary conditions. In particular, on the cylinder we have

$$Z_{a,b+c} = \chi_\epsilon(q) = \chi_{21}(q) + \chi_{31}(q), \tag{B15}$$
$$Z_{a,a+c} = \chi_\sigma(q) = \chi_{23}(q), \tag{B16}$$
$$Z_{a,a+b} = \chi_{\sigma^\dagger}(q) = \chi_{23}(q). \tag{B17}$$

The smallest scaling dimension in all of the above three cases is $\frac{1}{15}$ and we also have $2b^{a+c}_\sigma = 2b^{a+b}_\sigma = -b^{b+c}_\sigma = 2N\lambda^{-1}$.

The next interesting case is the free-free boundary condition with the cylinder partition function

$$Z_{f,f} = \chi_I(q) + \chi_\psi(q) + \chi_{\psi^\dagger}(q). \tag{B18}$$

It is not difficult to see that the operator with the smallest scaling dimension in this case is the energy operator with the conformal weight $\frac{2}{5}$. Note that in this case we have $b^a_\epsilon = b^b_\epsilon = b^c_\epsilon = N\lambda$. The last possible boundary conditions $(f, a)$ and $(f, a + b)$ can not be written with respect to the characters of the W symmetry. However, they were also studied in the literature and on the cylinder can be written with respect to the Virasoro characters as

$$Z_{f,a} = \chi_{(1,2)}(q) + \chi_{(4,2)}(q); \tag{B19}$$
$$Z_{f,b+c} = \chi_{(2,2)}(q) + \chi_{(3,2)}(q). \tag{B20}$$

In both cases the operator with the smallest scaling dimension is the energy operator and the behavior of the partition functions on the long cylinders are

$$Z_{f,a}(\tilde{q}) \to -\sqrt{3}N^2\lambda^2\tilde{q}^{\frac{2}{5}}; \tag{B21}$$
$$Z_{f,b+c}(\tilde{q}) \to \sqrt{3}N^2\tilde{q}^{\frac{2}{5}}. \tag{B22}$$

The above equations can be used to derive the sign of the Casimir force for all the boundary conditions studied here.